\documentclass[12pt]{article}

\makeatletter
\def\appendix{\par\clearpage
  \setcounter{section}{0}
  \setcounter{subsection}{0}
  \@addtoreset{equation}{section}
  \def\@sectname{Appendix~}
  \def\theequation{\thesection.\arabic{equation}}
  \def\thesection{\Alph{section}}}
\makeatother

\usepackage{amssymb}
\usepackage{graphicx}

\newcommand{\bea}{\begin{eqnarray}}
\newcommand{\eea}{\end{eqnarray}}

\begin{document}

\begin{titlepage}

\phantom{.}
\vspace{-3cm}

\hfill December 2004

\vskip 0.3cm

\centerline{\bf Analytical evolution of nucleon structure functions with 
power corrections}
\centerline{\bf at twist-4 and predictions for ultra-high energy 
neutrino-nucleon cross section}

\vskip 0.3cm

\centerline{R.~Fiore$^{a\dagger}$, L.L.~Jenkovszky$^{b\ddagger}$,
A.V.~Kotikov$^{c\star}$, F.~Paccanoni$^{d\ast}$,
A.~Papa$^{a\dagger}$, E.~Predazzi$^{e\$}$}

\vskip 0.2cm

\centerline{$^{a}$ \sl  Dipartimento di Fisica, Universit\`a della Calabria}
\centerline{\sl Istituto Nazionale di Fisica Nucleare, Gruppo collegato di Cosenza}
\centerline{\sl I-87036 Arcavacata di Rende, Cosenza, Italy}

\centerline{$^{b}$ \sl  Bogolyubov Institute for Theoretical Physics}
\centerline{\sl Academy of Sciences of Ukraine}
\centerline{\sl UA-03143 Kiev, Ukraine}

\centerline{$^{c}$ \sl Bogolyubov Laboratory of Theoretical Physics}
\centerline{\sl Joint Institute for Nuclear Research}
\centerline{\sl RU-141980 Dubna, Russia}

\centerline{$^{d}$ \sl Dipartimento di Fisica, Universit\`a di Padova}
\centerline{\sl Istituto Nazionale di Fisica Nucleare, Sezione di Padova}
\centerline{\sl via F. Marzolo 8, I-35131 Padova, Italy}

\centerline{$^{e}$ \sl Dipartimento di Fisica Teorica, Universit\`a di Torino}
\centerline{\sl Istituto Nazionale di Fisica Nucleare, Sezione di Torino}
\centerline{\sl via P. Giuria 1, I-10125 Torino, Italy}

\vskip 0.1cm

\begin{abstract}
In this paper we present an analytic result for the evolution in $Q^2$
of the structure functions for the neutrino-nucleon interaction, valid
at twist-2 in the region of small values of the Bjorken $x$ variable and for 
soft non-perturbative input. In the special case of flat initial
conditions, we include in the calculation also the contribution of the twist-4 
gluon recombination corrections, whose effect in the evolution is explicitly 
determined. Finally, we estimate the resulting charged-current 
neutrino-nucleon total cross section and discuss its behavior at ultra-high 
energies. 
\end{abstract}

\vskip 0.1cm
\hrule

\vfill

$
\begin{array}{ll}
^{\dagger}\mbox{{\it e-mail address:}} &  \mbox{fiore,~papa@cs.infn.it} \\
^{\ddagger}\mbox{{\it e-mail address:}} & \mbox{jenk@bitp.kiev.ua} \\
^{\star}\mbox{{\it e-mail address:}} & \mbox{kotikov@thsun1.jinr.ru} \\
^{\ast}\mbox{{\it e-mail address:}} & \mbox{paccanoni@pd.infn.it} \\
^{\$}\mbox{{\it e-mail address:}} & \mbox{predazzi@to.infn.it}
\end{array}
$

\end{titlepage}
\eject
\newpage

\section{Introduction}
\label{sec:intro}

Saturation effects in the parton distributions for a nucleon, that
could finally lead to unitarization of the cross sections, have
been thoroughly discussed in the past. In the small Bjorken $x$ region
these effects can be accounted for by the introduction of non-linear 
terms in the evolution equation for the gluon density~\cite{GLR}.
The study of non-linear evolution equations began twenty years 
ago and gave rise, in the following years, to different approaches to the 
small-$x$ region~\cite{MQ,BBS}. More recently, generalizations of non-linear 
evolution equations have been proposed with different
physical motivations~\cite{SCREEN,JM,RENO} and screening effects
have been incorporated in the theoretical framework in different
ways~\cite{GBW,BAL,KOV}. 

The approaches quoted above
find interesting applications in the interaction of ultra-high
energy neutrinos with nucleons and
nuclei~\cite{BACH,JALM,KKK,MVM}. While there is no general
consensus on the importance of screening effects, one expects,
at any rate, that the linear QCD evolution of parton
distribution functions will be tamed in the very small-$x$ region.
Predictions for the cross section of the neutrino-nucleon
interaction based on DGLAP~\cite{GL,DGLAP} or BFKL~\cite{power}
equations show a power increase with energy~\cite{KR,RG,KMS} that
will finally violate the Froissart bound. We notice that the
problem these papers address is rather intricate, since it requires
a complete knowledge of nucleon structure functions in both
variables $x$ and $Q^2$. At ultra-high energies the integrals,
giving the cross section in terms of structure functions,
cover the whole permissible range of values for these
variables and explore extreme regions of the $(Q^2,x)$ phase space,
where non-accelerator data exist. Moreover, the mathematical
complexity inherent in the solution of non-linear evolution
equations may conceal the physical essence of the problem. 

Simplifications are well possible if we limit ourselves to
the small-$x$ region only, with a warning about the consequences
on the value of the integrals giving the total cross section. For
example, a power-like behavior in $x$ of the parton distributions
is a simple solution of the DGLAP dynamics at next-to-leading
order on the region $x<0.1$ and for large $Q^2$ values, where the hard initial
condition $x^{-\lambda} \gg $ constant can be applied for parton distribution 
functions~\cite{KR}. Due to the asymptotic
behavior of the DGLAP evolution~\cite{GL}, small-$x$ data can also
be interpreted in terms of the ``double asymptotic
scaling"~\cite{ADR,BF,MSW} that provides an explicit solution to the
problem. The straightforward application of these approximation
schemes to the non-linear evolution is however questionable.
Moreover, it would be rather difficult to estimate the error
induced on the cross section. 

In a series of papers~\cite{WZ,WZA} a new set of evolution equations was
suggested that includes parton recombination. These new equations
are derived in the leading logarithmic $(Q^2)$ approximation and
differ from the traditional ones~\cite{GLR,MQ}, that rely on the
double logarithmic approximation (DLLA). DLLA means that only those
terms in the splitting functions that generate large logarithms in
$x$ are important. In other words, in the DLLA a diagram consists of
gluon ladders and any transition from gluon to quark is suppressed
in this approximation. It becomes difficult however to reconcile
small-$x$ approximations with the twist-4 gluon recombination
corrections of Ref.~\cite{WZA}. As emphasized in Ref.~\cite{JBA},
the twist-4 coefficient function, driven by a two-particle gluon
distribution, cannot be simplified by using DLLA or other
small-$x$ approximations. 

There is, however, the possibility to replace, at small $x$, the convolution 
of two functions by a simple product. The method, introduced in
Refs.~\cite{LY,M79,AVK,AKP}, allows a correct treatment of
the non-singular part of parton distributions and has found
numerous applications~\cite{AKP,FJKP,IKP}. This approach, to be
described at length in the following, has been applied also to the
evaluation of the contributions from higher-twist operators of the
Wilson operator product expansion~\cite{IKP}. The accuracy of this
method, when applied to the modified DGLAP equations of
Ref.~\cite{WZA}, can be verified a posteriori with a suitable
computer code. 

In this paper we study the
contributions of the twist-4 gluon recombination corrections to a
previous twist-2 calculation~\cite{FJKP}. In Ref.~\cite{FJKP} we
estimated the ultra-high energy neutrino-nucleon cross section in
the approach of Ref.~\cite{AKP}. The corrections introduced by the
presence of non-linear terms in the evolution equation, as given
in Ref.~\cite{WZA}, will change sensibly our previous estimate. 

In the next Section we will discuss the non-perturbative
input and present an analytical form for the $Q^2$ evolution. Some
restrictions on the value of the parameters of the input
distributions, present in Ref.~\cite{FJKP}, are relaxed in this new
formulation. However, gluon recombination corrections will be estimated
in the particular case of starting flat initial conditions. 
In Sections~\ref{sec:sigma} results are presented for the structure 
function $F_2^{\nu N}(x,Q^2)$ and for the charged-current neutrino-nucleon total
cross section. The asymptotic behavior of this cross 
section will be also discussed. In the Appendices the proof of the relevant
analytical results is presented.

\section{$Q^2$ evolution}
\label{sec:evolution}

As in our previous paper~\cite{FJKP}, we choose a soft
non-perturbative input based on analyses of the
nucleon structure functions~\cite{ELBA,DGJ,Rivista,DJLP,CAL,JLP}.
If we denote by $f_q(x,Q^2)$ the sea quark distribution
$xS(x,Q^2)$ and by $f_g(x,Q^2)$ the gluon distribution
$xG(x,Q^2)$, that is if we put
\begin{equation}
f_q(x,Q^2)\equiv xS(x,Q^2),\;\;\;\;\;f_g(x,Q^2)\equiv xG(x,Q^2)\;,
\label{e1}
\end{equation}
our soft non-perturbative input can be written in the form
\begin{eqnarray}
f_a (x,Q^2_0) = \left[ A_a + B_a \ln\left(\frac{1}{x}\right)
 \right](1-x)^{\nu(Q^2_0)} ~~~~~(a=q,g)\;, \label{e2}
\end{eqnarray}
where  $A_a$, $B_a$ and $\nu(Q^2_0)$ are unknown parameters to be
determined from data. Throughout this calculation at small $x$ we
will ignore the non-singlet quark component and limit ourselves to
the leading order (LO) of perturbation theory. 

The factor $(1-x)^{\nu(Q^2_0)}$ has been treated in
detail in Ref.~\cite{FJKP} and we neglect it in the following. At the end we
will take it into account by multiplying the resulting structure
function, $F_2^{\nu N}(x,Q^2)=f_q(x,Q^2)$ in the case of neutrino-nucleon
DIS, by an effective large-$x$ behavior $(1-x)^\nu$, with constant $\nu$.
Furthermore, as in Ref.~\cite{FJKP}, we define
\begin{displaymath}
t=\ln\left[\frac{\alpha_s(Q^2_0)}{\alpha_s(Q^2)}\right]=
\ln\left[\frac{\ln(Q^2/\Lambda^2_{LO})}
{\ln(Q^2_0/\Lambda^2_{LO})}\right],
\end{displaymath}
and the Ball-Forte scaling variables
\begin{equation}
\sigma=2\sqrt{-\hat d_{gg} t \ln(1/x)}\;, 
\hspace{2cm}\rho=\sqrt{\frac{-\hat d_{gg}t}{\ln(1/x)}}=\frac{\sigma}{2\ln(1/x)}\;,
\label{e3}
\end{equation}
where $\hat d_{gg} = -12/\beta_0$ and $\beta_0=11-2f/3$, with $f$ the number 
of flavors, is the LO coefficient of the QCD $\beta$-function (in units of 
$-16\pi^2$). For brevity, we introduce the notation 
$\overline d_{+}(1) = 1+20f/(27\beta_0)$, $d_{-}(1) = 16f/(27\beta_0)$. At the LO,
$\alpha_s(Q^2)=4\pi/(\beta_0\,\ln (Q^2/\Lambda_{LO}^2))$.

\subsection{DGLAP evolution}
\label{subsec: evolution}

In Ref.~\cite{FJKP} the results of Ref.~\cite{AKP} were used in order to
obtain an approximate evolution, in LO perturbation
theory, under the conditions $A_a\gg B_a$ (here and in the following the index
$a$ stands for $q$ or $g$), so that no interference appears in the $Q^2$ evolution 
of the coefficients multiplying the different powers of the logarithm. Since here we
relax these conditions, it is interesting to present the new
expressions for $f_q$ and $f_g$. 

The method of solution adopted in Ref.~\cite{AKP} can be summarized as follows.
Starting from the exact solution in the moment space, the anomalous
dimensions and  the coefficient functions are expanded in the
neighborhood of $n=1$. The singular part, when $n\to 1$, leads to
Bessel functions but, in order to achieve the accuracy $O(\rho)$, also
the regular part must be properly taken into account in the inverse
Mellin transform.

By analogy with Ref.~\cite{AKP}, it is possible to obtain the small-$x$ asymptotic
results for parton distribution functions (PDFs) and the $F_2$ structure function
at LO of perturbation theory by setting 
\begin{equation}
f_a(x,Q^2) = f_a^+(x,Q^2) + f_a^-(x,Q^2)\;, \label{e4}
\end{equation}
where
\begin{eqnarray}
f_a^-(x,Q^2)&=& \biggl[ A_a^- + B_a^- \ln\left(\frac{1}{x}\right)
\biggr] \cdot e^{- d_{-}(1) t} ~+~O(x)\;,
 \label{e5}\\
f_q^+(x,Q^2)&=& \biggl[ A_q^{+}\, \rho \, I_1(\sigma) + B_q^{+}
I_0(\sigma) \biggr] \cdot e^{-\overline d_{+}(1) t}  \cdot
\Bigl(1~+~O(\rho)\Bigr)\;, \label{e6} \\
f_g^+(x,Q^2)&=& \biggl[ A_g^{+}  I_0(\sigma) + B_g^{+}
\frac{1}{\rho} I_1(\sigma) \biggr] \cdot e^{-\overline d_{+}(1) t}
 \cdot
\Bigl(1~+~O(\rho)\Bigr)\;. \label{e7}
\end{eqnarray}
Here, $I_n(z)$ are modified Bessel functions and the coefficients
$A_a^{\pm}$ and $B_a^{\pm}$ are defined as
\begin{eqnarray}
B_q^{-} &=& B_q\;, \nonumber \\
B_g^{-} &=& -\frac{4}{9} B_q\;, \nonumber \\
A_q^{-} &=& A_q  ~-~ \frac{f}{9} \left( B_g + \frac{4}{9} B_q
\right)\;, \nonumber \\
A_g^{-} &=&  -\frac{4}{9} A_q - \frac{2}{27} \left(
\left(1-\frac{7f}{27} \right) B_q - \frac{2f}{3} B_g \right)\;,
\nonumber \\
B_g^{+} &=& B_g + \frac{4}{9} B_q~, \nonumber \\
B_q^{+} &=& \frac{f}{9} \left(B_g + \frac{4}{9} B_q \right)\;,
\nonumber \\
A_g^{+} &=& A_g + \frac{4}{9} A_q ~+~ \frac{2}{27} \left(
\left(1-\frac{7f}{27} \right) B_q ~-~ \frac{2f}{3} B_g \right)\;,
\nonumber \\
A_q^{+} &=&  \frac{f}{9} \left( A_g + \frac{4}{9} A_q ~-~
\frac{1}{6} \left( 1+\frac{7f}{27} \right) B_g - \frac{4f}{243}
B_q \right)\;. \label{e8}
\end{eqnarray}

\subsection{Shadowing and anti-shadowing corrections}
\label{subsec:shadow}

According to Ref.~\cite{WZA}, sea quark and gluon distributions are
modified by the introduction of gluon recombination as stated by
the following modified DGLAP equations:
\begin{eqnarray}
&&\frac{df_a^{full}(x,Q^2)}{d\ln Q^2} = \sum_{b=q,g}
P^{AP}_{ab}(x) \otimes
f_b^{full}(x,Q^2) \nonumber \\
&& + \frac{\alpha_s^2}{Q^2} \left[ K_1 \int^x_{x/2} \frac{dy}{y}
F_{ag}\left(\frac{x}{y}\right) \left({f_g^{full}(y,Q^2)}\right)^2
\right. \nonumber \\ & & \left. -K_2 \int_x^{1/2} \frac{dy}{y}
F_{ag}\left(\frac{x}{y}\right) \left({f_g^{full}(y,Q^2)}\right)^2
\right]\;. \label{e9}
\end{eqnarray}
Here $P^{AP}_{ab}(x)$ are the Altarelli-Parisi kernels, $\otimes$ stands for 
the Mellin convolution, defined as 
\begin{displaymath}
A(x) \otimes B(x) = \int^1_{x} \frac{dy}{y}
A\left(\frac{x}{y}\right) B(y) \;,
\end{displaymath}
and
\begin{eqnarray}
F_{gg}(x) &=& \frac{27}{64} (2-x) \Bigl( 99-136x+132x^2-64x^3+16
x^4\Bigr)\;, \label{e10} \\
F_{qg}(x) &=& \frac{x}{48} (2-x) \Bigl( 36-60x+49x^2-14x^3\Bigr)\;.
\label{e11}
\end{eqnarray}
In Ref.~\cite{WZA} it was set $K_1=K_2=K$ and from a fit of the HERA data the 
resulting value for $K$ in the evolution equations of Ref.~\cite{WZA} turned
out to be very small, $K=0.0014$. We will come back to this point in 
Subsection~\ref{subsec:fit}. The introduction of two
parameters, $K_1$ and $K_2$, allows a clearer check of the
importance of the anti-shadowing, with respect to the shadowing
contribution. Moreover, they give the possibility to relate models introduced in
Refs.~\cite{WZ,WZA} and \cite{GLR,MQ} (see Subsection~\ref{subsec:fit}). 

Since $K_1$ and $K_2$ are expected to be small numbers, the solution of 
Eq.~(\ref{e9}) can be written as
\begin{equation}
f_a^{full}(x,Q^2) = f_a(x,Q^2) + T_a(x,Q^2)\;, \label{e12}
\end{equation}
where
\begin{equation}
\frac{df_a(x,Q^2)}{d\ln Q^2} = \sum_{b=q,g} P^{AP}_{ab}(x)
\otimes f_b(x,Q^2) \label{e13}
\end{equation}
and
\begin{equation}
\frac{dT_a(x,Q^2)}{d\ln Q^2} = \sum_{b=q,g} P^{AP}_{ab}(x)
\otimes T_b(x,Q^2) + \alpha_s \, R_a(x,Q^2)\;, \label{e14}
\end{equation}
with
\begin{eqnarray}
R_a(x,Q^2) &=& \frac{\alpha_s}{Q^2} \left[ K_1 \int^x_{x/2}
\frac{dy}{y} F_{ag}\left(\frac{x}{y}\right)
\left({f_g(y,Q^2)}\right)^2 \right. \nonumber \\ &-&  \left. K_2
\int_x^{1/2} \frac{dy}{y} F_{ag}\left(\frac{x}{y}\right)
\left({f_g(y,Q^2)}\right)^2 \right] \;.\label{e15}
\end{eqnarray}
Finally, the sea quark distribution is
\begin{displaymath}
xS(x,Q^2) = f_q^{full}(x,Q^2)\;.
\end{displaymath}

\subsection{Small-$x$ solution of the complete equations}
\label{subsec:small-x}

The solution of Eq.~(\ref{e13}) with the boundary condition~(\ref{e2})
has been found already and is expressed in Eqs.~(\ref{e4})-(\ref{e7}). This result
has been obtained directly from the corresponding solution in the
moment space (see Ref.~\cite{AKP}). In order to simplify the solution
of the complete equations, we set in the following $B_q=B_g=0$.
Then our solution~(\ref{e5})-(\ref{e7}) assumes the simple
form
\begin{eqnarray}
f_a^-(x,Q^2)&=&  A_a^- \cdot e^{- d_{-}(1) t} ~+~O(x)\;,
 \label{u5}\\
f_q^+(x,Q^2)&=& A_q^{+}\, \rho \, I_1(\sigma) \cdot e^{-\overline
d_{+}(1) t}
 \cdot
\Bigl(1~+~O(\rho)\Bigr) \;, \label{u6} \\
f_g^+(x,Q^2)&=& A_g^{+}  I_0(\sigma)\cdot e^{-\overline d_{+}(1)
t} \cdot \Bigl(1~+~O(\rho)\Bigr) \;.\label{u7}
\end{eqnarray}
Equation~(\ref{e14}) can be rewritten in the moment space, using the Mellin 
transform defined as 
\[
M(n)= \int^{1}_0 \, dx \, x^{n-2} \, M(x)\;.
\]
In the leading logarithmic approximation we obtain
\begin{equation} \frac{dT_a(n,Q^2)}{dt} =
-\sum_{b=q,g} d_{ab} T_b(n,Q^2) + r_a(n,Q^2) \;, \label{e16}
\end{equation}
where $d_{ab}(n)=\gamma_{ab}^{(0)}(n)/(2\beta_0)$ is the ratio between
the anomalous dimension $\gamma_{ab}^{(0)}$ and twice $\beta_0$ and 
$r_a(n,Q^2)=4\pi R_a(n,Q^2)/\beta_0$, with $R_a(n,Q^2)$ the Mellin moment of 
$R_a(x,Q^2)$. The solution of Eq.~(\ref{e16}) can be written in the form
\begin{eqnarray}
T_a(n,Q^2)&=& T_a^+(n,Q^2)~+~ T_a^-(n,Q^2)\; , \nonumber \\
T_a^{\pm}(n,Q^2)&=& - \int^{\infty}_t dw ~ e^{d_{\pm}(w-t)}
\sum_{b=q,g} \epsilon^{\pm}_{ab}r_a(n,M^2) \;, \label{e17}
\end{eqnarray}
where
\begin{displaymath}
w=\ln
\left(\frac{\ln(M^2/\Lambda_{LO}^2)}{\ln(Q^2_0/\Lambda_{LO}^2)}\right)\;,
\end{displaymath}
$d_{\pm}(n)$ are the eigenvalues of the $d_{ab}$ matrix and
$\epsilon^{\pm}_{ab}(n)$ are related to the components of the eigenvectors
of the same matrix. Explicitly, we have
\begin{eqnarray}
T_q^{-}(n,Q^2) &=& - \int^{\infty}_t dw \, e^{d_{-}(w-t)}
\biggl[\eta r_q(n,M^2) +
\tilde \eta r_g(n,M^2) \biggr]\;, \nonumber \\
T_q^{+}(n,Q^2) &=& - \int^{\infty}_t dw \, e^{d_{+}(w-t)}
\biggl[(1-\eta ) r_q(n,M^2) -
\tilde \eta r_g(n,M^2) \biggr], \nonumber \\
T_g^{-}(n,Q^2) &=& - \int^{\infty}_t dw \, e^{d_{-}(w-t)}
\biggl[(1-\eta) r_g(n,M^2) +
\epsilon  r_q(n,M^2) \biggr]\;, \nonumber \\
T_g^{+}(n,Q^2) &=& - \int^{\infty}_t dw \, e^{d_{+}(w-t)}
\biggl[\eta r_g(n,M^2) - \epsilon r_q(n,M^2) \biggr]\;,
\label{e17b}
\end{eqnarray}
where $\epsilon^{\pm}_{ab}$ have been expressed in terms of
$\eta, \tilde \eta$ and $\epsilon$, which take the following values
for $n \to 1$:
\begin{equation} \eta =1-\frac{4f}{81} (n-1),
\hspace{1cm} \tilde \eta = -\frac{f}{9} (n-1),
\hspace{1cm} \epsilon = -\frac{4}{9} \;.\label{e18}
\end{equation}

The evaluation of the Mellin moments of $R_a(x,Q^2)$
requires particular care, since the integrals appearing in Eq.~(\ref{e15}) 
are not exact Mellin convolutions. If we make the position
\[
M(n|1/2) = \int^{1/2}_0 \, dx \, x^{n-2} \, M(x) \;,
\]
we have, in addition to the usual Mellin convolution 
\begin{equation}
\int^{1}_0 \, x^{n-2} \, dx \int^{1}_x \, \frac{dy}{y} \,
M_1\left(\frac{x}{y}\right)
M_2(y) = M_1(n)\,M_2(n)\;, 
\label{e19} 
\end{equation}
also the following one:
\begin{equation}
\int^{1}_0 \, x^{n-2} \, dx \int^{1/2}_{x/2} \, \frac{dy}{y} \,
\tilde M_1\left(\frac{x}{2y}\right) M_2(y) = 2^{n-1}\, \tilde
M_1(n)\,M_2(n|1/2) \;,
\label{e20}
\end{equation}
where the definition $\tilde M_1(x) \equiv M_1(2x)$ is used hereafter.
Now, in order to calculate the Mellin moments of $R_a(x,Q^2)$ in Eq.~(\ref{e15}), 
we should find a relation between the ``Mellin transforms" $f_2(n)$ and 
$f_2(n|1/2)$, defined as
\begin{eqnarray}
f_2(n)&=& \int^{1}_0 \, dx \, x^{n-2} \, f^2_g(x)\;, \label{e21} \\
f_2(n|1/2) &=& \int^{1/2}_0 \, dx \, x^{n-2} \, f^2_g(x)\;.
\label{e22}
\end{eqnarray}
This relation is
\begin{equation}
f_2(n|1/2)=f_2(n)+O((n-1)^2) \;,\label{e23}
\end{equation}
the proof being given in Appendix~A. 

Then, the contributions to the Mellin transform of $R_a(x,Q^2)$ which are
regular for $n \rightarrow 1$ assume the form (the upper index $(r)$, here
and in the following, stands for ``regular contribution'')
\begin{equation}
\Bigl[ \tilde F_{ag}^{(r)}(n) - F_{ag}^{(r)}(n) \Bigr] \, f_2(n)\;,
\label{e24}
\end{equation}
for the coefficient of $K_1$ and 
\begin{equation}
F_{ag}^{(r)}(n) \, f_2(n)\;, \label{e25}
\end{equation}
for the coefficient of $K_2$, where (see also Eqs.~(\ref{e19}) and (\ref{e20}))
\begin{eqnarray}
F_{ag}^{(r)}(n) &=& \int^1_0 dx \, x^{n-2} F_{ag}^{(r)}(x)\;, \nonumber \\
\tilde F_{ag}^{(r)}(n) &=& \int^1_0 dx \, x^{n-2} \tilde
F_{ag}^{(r)}(x) = \int^1_0 dx \, x^{n-2} F_{ag}^{(r)}(2x)\;,
\label{e27}
\end{eqnarray}
since by definition $F_{ag}^{(r)}(x) \equiv \tilde F_{ag}^{(r)}(x/2)$.

Note that $F_{qg}^{(r)}(x)=F_{qg}(x)$, because the Mellin moments
of $F_{qg}(x)$ have no singular part, and
\begin{equation}
F_{gg}^{(r)}(x) = \frac{27}{64} \Bigl[-99x -4x(2-x) \Bigl(
34-33x+16x^2-4x^3\Bigr)\Bigr]\;. \label{e28}
\end{equation}
By performing the Mellin integrals in Eqs.~(\ref{e27}), we find
\begin{eqnarray}
&& F_{qg}^{(r)}(n=1) = \frac{1813}{2880}\;, \nonumber \\
&& \tilde F_{qg}^{(r)}(n=1) = \frac{131}{180}\;, \nonumber \\
&& F_{gg}^{(r)}(n=1) = -\frac{31977}{320}\;, \nonumber \\
&& \tilde F_{gg}^{(r)}(n=1) = -\frac{23877}{160}\;.
\label{e29}
\end{eqnarray}
In this way we obtain from Eqs.~(\ref{e15}), (\ref{e24}) and (\ref{e25})
\begin{eqnarray} R_{q}(n\to 1,M^2) &=&
\left[ \frac{283}{2880}\, K_1 - \frac{1813}{2880}\, K_2 \right] \,
\frac{\alpha_s(M^2)}{M^2} \, f_2(n\to 1,M^2)\;, \label{e30} \\
R_{g}(n\to 1,M^2) &=&  9 \, \left[ \left(\frac{297}{32}\ln 2 -
\frac{1753}{320}\right)\, K_1 - \left(\frac{297}{32}\frac{1}{n-1}
-\frac{3553}{320}\right) \, K_2 \right] \nonumber \\
&\times& \frac{\alpha_s(M^2)}{M^2} \, f_2(n\to 1, M^2) \label{e31}
\end{eqnarray}
and, from Eqs.~(\ref{e17b}),
\begin{equation}
T_a^{\pm}(n\to 1,Q^2) = -\int^{\infty}_{Q^2} \!\! \frac{dM^2}{(M^2)^2} 
C^{\pm}_{ag}(n\to 1) f_2(n\to1,M^2)
\frac{{\left(\alpha_s(Q^2)\right)}^{d_{\pm}(n\to 1)}}{{\left(
\alpha_s(M^2)\right)}^{d_{\pm}(n\to 1)-2}} \,.\label{e32}
\end{equation}
Here $ C^{\pm}_{ag}(n \to 1)$ are the ``coefficient functions'' of the
twist-4 corrections, because 
\bea
R_a(n \to 1,Q^2) &=&  \frac{\alpha_s(Q^2)}{Q^2} C_{ag}(n \to 1)
f_2(n \to 1,Q^2), \nonumber \\
C_{ag}(n \to 1) &=& C^{+}_{ag}(n \to 1) + C^{-}_{ag}(n \to 1)\;,
\label{dop}
\eea
i.e. they are the coefficients in front of the Mellin moments
of the function $f_g^2(x)$. They can be written as
\begin{equation}
C^{\pm}_{ag}(n\to 1) = C^{1,\pm}_{ag}\,K_1-C^{2,\pm}_{ag}\,K_2
\label{e33}
\end{equation}
and the definitions $C^{i,\pm}_{ag}(n\to 1)\equiv C^{i,\pm}_{ag}$
$(i=1,2)$ have been introduced for the sake of simplicity. The
coefficients $C^{i,\pm}_{ag}$ are given in Appendix~B. 

The solution~(\ref{e32}) in the moment space can be
transformed to the $x$-space. Note that the product of moments
in Eq.~(\ref{e32}) leads to the convolution
\begin{equation}
M_1(n)\cdot M_2(n) ~\stackrel{M^{-1}}{\to}~ \int^1_x \frac{dy}{y}
\, M_1\left(\frac{x}{y}\right) \, M_2(y) \equiv M_1(x) \otimes M_2(x) \label{e34}
\end{equation}
in the $x$-space.
As in the case of the moment space, $T_a(x,Q^2)$ can be
represented as the combination of the ``$+$'' and ``$-$''
components:
\begin{equation}
T_a(x,Q^2) = T_a^+(x,Q^2) + T_a^-(x,Q^2)\;, \label{e35}
\end{equation}
which can be obtained from the corresponding components in
Eq.~(\ref{e32}) by an inverse Mellin transformation.

As for the ``$-$'' component, we note that the value $d_{-}(n)$ does not contain any
singularity for $n\to1$, hence (hereafter $v=w-t$)
\begin{equation}
e^{d_{-}(n)v} \approx e^{d_{-}(1)v}~~ \stackrel{M^{-1}}{\to}~~
e^{d_{-}(1)v} \delta(1-x) \label{e36}
\end{equation}
(here $\delta(1-x)$ is the Dirac $\delta$-function), so that
\begin{equation}
 C^{-}_{ag} \, e^{d_{-}(1)(w-t)} \, f_2(n\to 1,M^2)
~~\stackrel{M^{-1}}{\to} ~~
 C^{-}_{ag} \, e^{d_{-}(1)(w-t)} \, f^2_g(x,M^2)\;.
\label{e37}
\end{equation}
As for the ``$+$'' component, we have $d_+(n)=\hat d_+/(n-1) +
\overline d_+(n)$, with $\hat d_+<0$, hence
\begin{eqnarray}
&&\frac{1}{n-1}e^{\hat d_+v/(n-1)} = \sum_{k=0}^{\infty}
\frac{1}{k!} \, \frac{(\hat d_+v)^k}{(n-1)^{k+1}} \nonumber \\
\stackrel{M^{-1}}{\to} && \sum_{k=0}^{\infty}
\frac{1}{(k!)^2} \, (\hat d_+v)^k {\left(\ln
\frac{1}{x}\right)}^{k} = J_0(\tilde \sigma)\;,  \label{e38} \\
&&e^{\hat d_+v/(n-1)} = \sum_{k=0}^{\infty}
\frac{1}{k!} \, \frac{(\hat d_+v)^k}{(n-1)^{k}} \nonumber \\
\stackrel{M^{-1}}{\to} && \delta(1-x) - \sum_{k=0}^{\infty}
\frac{1}{k!(k-1)!} \, (\hat d_+v)^k {\left(\ln
\frac{1}{x}\right)}^{k-1} \nonumber \\ && = \delta(1-x) -\tilde
\rho J_1(\tilde \sigma)\;, \label{e39}
\end{eqnarray}
where 
\begin{eqnarray}
{\hat{\rho}}&=&\frac{{\hat{\sigma}}}{2\ln(1/x)}, \hspace{1cm}
{\hat{\sigma}} = \sigma \mbox{ with } t \to w\;, \label{e40} \\
{\tilde{\rho}}&=&\frac{{\tilde{\sigma}}}{2\ln(1/x)}, \hspace{1cm}
{\tilde{\sigma}} = \sigma \mbox{ with } t \to (w-t) \label{e41}
\end{eqnarray}
and $w=t$ when  $Q^2 \to M^2$, i.e.
\begin{eqnarray}
w&=&\ln\left(\frac{\alpha_s(Q^2_0)}{\alpha_s(M^2)}\right)=
\ln\left(\frac{\ln(M^2/\Lambda^2_{LO})}
{\ln(Q^2_0/\Lambda^2_{LO})}\right)~, \label{e42} \\
w-t&=&\ln\left(\frac{\alpha_s(Q^2)}{\alpha_s(M^2)}\right)=
\ln\left(\frac{\ln(M^2/\Lambda^2_{LO})}
{\ln(Q^2/\Lambda^2_{LO})}\right)\;. \label{e43}
\end{eqnarray}
As a consequence
\begin{eqnarray}
&& C^{+}_{qg}(n=1) \, e^{\hat d_{+}(w-t)/(n-1)}
\, e^{\overline d_{+}(1)(w-t)} \, f_2(n\to 1,M^2) \nonumber \\
\stackrel{M^{-1}}{\to} &&
 C^{+}_{qg}(n= 1) \, e^{\overline d_{+}(1)(w-t)} \, \tilde F_1\;,
\label{e44}\\
&& \Bigl[\hat C^{+}_{gg}\frac{1}{n-1} + \overline C^{+}_{gg}(n=1)
\Bigr] \, e^{\hat d_{+}(w-t)/(n-1)}
\, e^{\overline d_{+}(1)(w-t)} \, f_2(n\to 1,M^2) \nonumber \\
\stackrel{M^{-1}}{\to} && \Bigl[\hat C^{+}_{gg} \, \tilde F_2
+ \overline C^{+}_{gg}(n=1) \, \tilde F_1 \Bigr]\,e^{\overline
d_{+}(1)(w-t)}\;, \label{e45}
\end{eqnarray}
where
\begin{eqnarray}
{\tilde{F_1}} &=& - \int_x^{1} \frac{dy}{y} \, \Bigl[\delta(1-y) -
\tilde \rho(y) J_1(\tilde \sigma(y))\Bigr] \,
f^2_g(x/y,M^2) \nonumber \\
&\equiv& \Bigl[\delta(1-x) - {\tilde{\rho}} J_1({\tilde{\sigma}})
\Bigl] \otimes
f^2_g(x,M^2)\;, \label{e46} \\
\tilde F_2 &=& \int_x^{1} \frac{dy}{y} \, J_0(\tilde \sigma(y)) \,
f^2_g(x/y,M^2) \equiv J_0({\tilde{\sigma}}) \otimes f^2_g(x,M^2)\;.
\label{e47}
\end{eqnarray}
Thus,
\begin{eqnarray}
T_a^-(x,Q^2)&=& -\int^{\infty}_{Q^2} \, \frac{dM^2}{(M^2)^2}
\Bigl( K_1 \cdot C^{1,-}_{ag} - K_2 \cdot C^{2,-}_{ag} \Bigr) \,
\left({f_g(y,M^2)}\right)^2 \nonumber \\  &\times&
\frac{{\left(\alpha_s(Q^2)\right)}^{d_{-}(1)}}{{\left(
\alpha_s(M^2)\right)}^{d_{-}(1)-2}}\;,
 \label{e48}\\
T_q^+(x,Q^2)&=& -\int^{\infty}_{Q^2} \, \frac{dM^2}{(M^2)^2}
\Bigl( K_1 \cdot C^{1,+}_{qg} - K_2 \cdot C^{2,+}_{qg} \Bigr) \,
{\tilde{F_1}} \nonumber \\  & \times &
\frac{{\left(\alpha_s(Q^2)\right)}^{{\overline{d}}_{+}(1)}}{{\left(
\alpha_s(M^2)\right)}^{{\overline{d}}_{+}(1)-2}}\;,
\label{e49} \\
T_g^+(x,Q^2)&=& -\int^{\infty}_{Q^2} \, \frac{dM^2}{(M^2)^2}
\Bigl( K_1 \cdot C^{1,+}_{gg} \, {\tilde{F_1}} - K_2 \, \Bigl[\hat
C^{2,+}_{gg} \, \tilde F_2 + \overline C^{2,+}_{gg} \, \tilde F_1
\Bigr]  \, \Bigr) \, \nonumber \\ & \times &
\frac{{\left(\alpha_s(Q^2)\right)}^{{\overline{d}}_{+}(1)}}{{\left(
\alpha_s(M^2)\right)}^{{\overline{d}}_{+}(1)-2}}\;.
 \label{e50}
\end{eqnarray}
The above formulas complete our calculation of the gluon
recombination terms.

\subsection{Evaluation of $\tilde{F_1}$ and $\tilde{F_2}$ at
O($\rho$)}
\label{subsec:tilde}

The functions ${\tilde{F_1}}$ and ${\tilde{F_2}}$ can be evaluated
approximately. Note that we can write
\begin{eqnarray}
{\tilde{F_1}} &=& \tilde F^{--}_1 + 2  \tilde F^{+-}_1 +
 \tilde F^{++}_1\;, \nonumber \\
{\tilde{F_2}} &=& \tilde F^{--}_2  + 2  \tilde F^{+-}_2 +
 \tilde F^{++}_2\;,
\label{e51}
\end{eqnarray}
according to the decomposition (see Eq.~(\ref{e4}))
\begin{displaymath}
f_g^2(x,M^2) = (f^{-}_g(x,M^2))^2 +2 f^{+}_g(x,M^2)f^{-}_g(x,M^2)
+ (f^{+}_g(x,M^2))^2 \;.
\end{displaymath}
The details of the complete calculation, at order $O(\rho)$ and with 
$B_q=B_g=0$, will be given in Appendix~C. Here we present only the results.

For the ``$- -$'' component we find
\begin{eqnarray}
\tilde F^{--}_1 &=& (A_g^-)^2 J_0({\tilde{\sigma}}) \cdot e^{-
2d_{-}(1) w} \Bigl(1~+~O(\rho)\Bigr)\;, \label{e52} \\
\tilde F^{--}_2 &=& (A_g^-)^2 \frac{1}{\tilde \rho}
J_1({\tilde{\sigma}}) \cdot e^{- 2d_{-}(1) w}
\Bigl(1~+~O(\rho)\Bigr)\;, \label{e53}
\end{eqnarray}
while the result for the ``$+ -$'' component is
\begin{eqnarray}
\tilde F^{+-}_1 &\equiv& - {\tilde{\rho}} J_1({\tilde{\sigma}})
\otimes  \Bigl(f^{+}_g(x,M^2) f^{-}_g(x,M^2)\Bigr) \nonumber \\
&=& A_g^- A_g^{+} I_0(\sigma) \cdot e^{- (d_{-}(1)+\overline
d_{+}(1)) w} \Bigl(1~+~O(\rho)\Bigr)\;,
 \label{e54} \\
\tilde F^{+-}_2 &\equiv& J_0({\tilde{\sigma}})
\otimes  \Bigl(f^{+}_g(x,M^2)f^{-}_g(x,M^2)\Bigr) \nonumber \\
&=& A_g^- A_g^{+} \frac{1}{\rho} I_1(\sigma) \cdot e^{-
(d_{-}(1)+\overline d_{+}(1)) w} \Bigl(1~+~O(\rho^2)\Bigr)\;.
 \label{e55}
\end{eqnarray}
The term corresponding to the component ``$+ +$'' requires a more
involved treatment. As shown in Appendix~C, one obtains 
\begin{eqnarray}
\tilde F^{++}_1 &\equiv& \Bigl[\delta(1-x) - {\tilde{\rho}}
J_1({\tilde{\sigma}}) \Bigr]
\otimes  \Bigl(f^{+}_g(x,M^2)\Bigr)^2 \nonumber \\
&=& ( A_g^{+})^2 \,I_0(zy^{1/2})I_0(\bar{z}y^{1/2}) \cdot e^{-2
\overline d_{+}(1) w}
 \cdot
\Bigl(1~+~O(\rho)\Bigr) \label{e56}
\end{eqnarray}
and
\begin{eqnarray}
\tilde F^{++}_2 &\equiv& J_0({\tilde{\sigma}})
\otimes  \Bigl(f^{+}_g(x,M^2)\Bigr)^2 \nonumber \\
&=& ( A_g^{+})^2 \,\frac{1}{2\sqrt{\alpha\beta}} y^{1/2}\left[ z
I_0(\bar{z}y^{1/2})I_1(zy^{1/2}) \right. \nonumber \\ & - & \left.
\bar{z} I_0(z y^{1/2}) I_1(\bar{z}y^{1/2}) \right] \cdot e^{-2
\overline d_{+}(1) w} \cdot \Bigl(1~+~O(\rho^2)\Bigr)\;, \label{e57}
\end{eqnarray}
where $y=-\log(x)$, $z=\alpha^{1/2}+i \left|\beta^{1/2} \right|$,
with $\alpha=-\hat{d}_+ (3w+t)$ and $\beta=-\hat{d}_+ (t-w)$, and $\bar{z}$
is the complex conjugate of $z$.

These approximations do not make the integrations in Eqs.~(\ref{e49})-(\ref{e50}) 
avoidable, but greatly simplify the analytic structure of the answer.

\section{The neutrino-nucleon cross section}
\label{sec:sigma}

\subsection{The fitting procedure}
\label{subsec:fit}

We consider the parton distribution functions from the ZEUS 
Collaboration~\cite{ZEUS} 
in the region $2.5$ GeV$^2 \leq Q^2 \leq 20$ GeV$^2$, for values of $x$ in the 
range $10^{-4} \leq x \leq 5\times 10^{-3}$, where the ZEUS NLO fit favorably 
compares with existing HERA data. From the parton
distributions it is possible to reconstruct the isoscalar structure
function $2xF_1$ for neutrino-nucleon scattering
\begin{displaymath}
2xF_1^{\nu N}(x)\simeq 2xF_1^{\bar{\nu} N}(x)
\end{displaymath}
\begin{equation}
\simeq xu(x)+x\bar{u}(x)+xd(x)+x\bar{d}(x)+2xs(x)+2xc(x)+\ldots
\label{nu1}
\end{equation}
where $\ldots$ stands for $b$ and $t$ quarks and we have assumed
$s=\bar{s}$ and $c=\bar{c}$. At the LO $F_2(x)=2x F_1(x)$ and the
difference between the two structure functions decreases when $Q^2$
increases. Parton distribution functions are needed at any rate;
the knowledge of the experimental $F_2^{\ell N}$ must in fact be
supplemented by PDFs since
\begin{displaymath}
F_2^{\nu N}=\frac{18}{5}F_2^{\ell N}+\frac{6}{5}(xs-xc)+\ldots
\end{displaymath}

For the $Q^2$ values under consideration, there are no CCFR
data~\cite{CCFR} below $x=0.0125$ and the comparison between our fit and 
$F_2$ measurements in the process $(\nu^{\mu}+\bar{\nu}^{\mu})+
\mbox{nucleon} \rightarrow (\mu^-+\mu^+)+X$ is impossible. A
single high energy HERA data point~\cite{H1COLL} for the process
$e^-p\rightarrow \nu_eX$, with $p_{\bot}> 25$ GeV, gives
$\sigma=55\pm 15$ pb at $\sqrt{s}=296$ GeV. This means for the
neutrino-nucleon cross section a value of $(2.0\pm 0.55)\times
10^{-34}$ cm$^2$ at $\sqrt{s}=306.4$ GeV. 

The possibility to test other saturation models, such as that of 
Ref.~\cite{GLR}, relies on the presence of two different coefficients $K_1$ 
and $K_2$ for shadowing and antishadowing contributions. In the following, 
we set $K_1=K_2=K$, as in Ref.~\cite{WZA}, thus reducing the number of parameters
in our approach and the related errors. Then, the free parameters become
$\Lambda_{LO},\,f,\,Q_0,\,A_q,\,A_g,\,K$. 
We choose to fix $f=4$, as in Ref.~\cite{IKP}, and $\Lambda_{LO}=0.19$ GeV. 
The remaining parameters should be 
determined by a fit. There is one more parameter to be considered,
namely the ``mean power" $\nu$ of the factor $(1-x)$ present in
$F_2^{\nu N}(x,Q^2)$. The term $(1-x)^\nu$ will be considered, for the sake of
simplicity, only at the end of the calculation and will not
evolve in our model. This will surely affect the result, but, since
we are interested in the total cross section at very high energy,
and hence at $\langle x\rangle$ very small, the error should not influence too
much the result. 

The values of the parameters $Q_0$ and $K$ can be estimated from other sources. 
In the paper by W.~Zhu {\it et al.}~\cite{WZA} a fit to the HERA small-$x$ data for
$F_2^{ep}$ has been performed starting from GRV-like~\cite{GRV} input
distributions at $Q_0^2=0.34$ GeV$^2$. MD-DGLAP evolution equations~\cite{WZA}
determine new parameters for the sea quark distributions at
$Q_0^2$, with respect to GRV98LO~\cite{GRV}, and give for the non-linear
coefficient $K$ the value $K=0.0014$. 

On the other hand, the fit of Ref.~\cite{IKP} imposes a constraint on the parameter
$Q_0$. This constraint originates from the conditions of
applicability of the ``generalized double asymptotic scaling", that
gives a satisfactory description of the experimental data in a
region $Q^2> Q^2_{cut}$, where $Q^2_{cut}$ is a cutoff larger than
$Q_0^2$. For small $x$ and large $Q^2$ values it is reasonable to
neglect the valence quarks and the disagreement at small $Q^2$ is
a consequence of this approximation. The value of $Q_0^2$ in this
approach, that we follow in our paper, turns out to be rather
small and, for the LO fit, it is approximately $Q_0^2
\sim 0.3\div 0.4$ GeV$^2$. This value of $Q_0^2$ has been obtained
by fitting HERA data with higher-twist contribution evaluated in
the renormalon model~\cite{REN}. 

At this point it can be useful to remember that the purpose of this calculation 
is to reproduce the neutrino-nucleon cross section at very high energy,
where presumably valence quarks do not contribute. Looking at Ref.~\cite{KMS}, 
in particular at the Figure~2 in this paper, one
sees that the valence quark contribution to the total $\nu N$
charged-current cross section is larger than the $u$ and $d$ sea
quarks contribution up to the energy $E_{\nu}\sim 10^5$ GeV.
Hence we cannot rely on the H1 Collaboration data point at
$E_{\nu}\sim 5\times 10^4$ GeV and we expect that our model can
be trusted only at higher energies.

\subsection{Results}
\label{subsec:res}

We have reconstructed the structure function $F_2^{\nu N}(x,Q^2)$
from the ZEUS PDFs~\cite{ZEUS} and
performed a fit to the asymptotic formula for $F_2^{\nu N}(x,Q^2)
=f_q^{full}(x,Q^2)$, where $f_q^{full}(x,Q^2)$ has been defined in 
Eq.~(\ref{e12}) and the ingredients for calculating it have been given in 
Section~\ref{sec:evolution}. 
Basing ourselves on the arguments of Ref.~\cite{IKP}, we impose
an upper bound on $Q_0^2$, $Q_0^2 \leq 0.45$ GeV$^2$, and consider separately
two possible values for the mean power $\nu$ of the factor $(1-x)$: $\nu=4$ or
$\nu=5$. The lowest $\chi^2$/d.o.f for this constrained fit has been obtained
having chosen $Q_0^2 = 0.45$ GeV$^2$ and $\nu=4$, with the result 
($\chi^2$/d.o.f.=0.553)
\begin{eqnarray}
A_q &=& 1.002(39) \nonumber \\  
A_g &=& 0.565(29) \nonumber \\
K &=& 0.0130(49) \;. \label{fit}
\end{eqnarray}

There are several reasons why the parameter $K$ in~(\ref{fit}) is much larger than
the corresponding result in Ref.~\cite{WZA}. First of all, the change from the
leptonic structure function of Ref.~\cite{WZA} to the neutrino one requires a factor
near to 18/5 that affects also the coefficient $K$. In addition, our fit
shows that there is a strong correlation between the starting value of $Q^2$,
i.e. $Q_0^2$, and the value of $K$, which turns out to increase with $Q_0^2$. 
These reasons suffice to say that, within errors, our result is compatible with 
that of Ref.~\cite{WZA}.

In order to test our solution, we have calculated $F_2^{\nu N}(x,Q^2)$ from 
the ZEUS PDFs~\cite{ZEUS} at $Q^2=50$ GeV$^2$ and compared with 
our theoretical result. The percentage error at $x=10^{-4}$ is nearly 2\% and is
of the same order of magnitude in the whole range of $x$: $ 10^{-4}
\leq x\leq 5\times 10^{-3}$ (see Figure~1). 

\begin{figure}[tb]
\centering
\includegraphics[width=\textwidth]{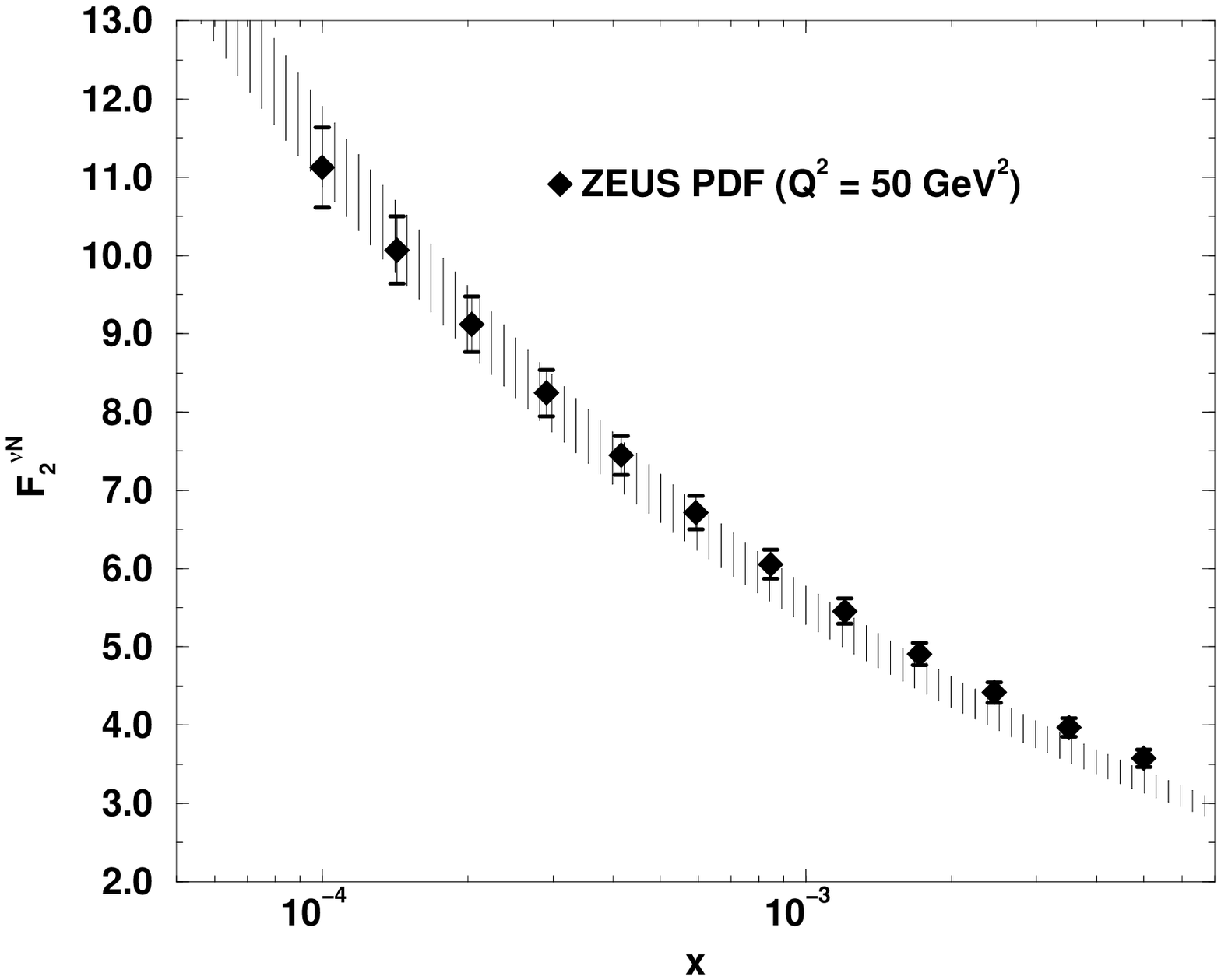}
\caption[]{\small Comparison at $Q^2=50$ GeV$^2$ between $F_2^{\nu N}(x,Q^2)$ 
as obtained from the ZEUS PDFs~\cite{ZEUS} and from our theoretical calculation (the
vertical bars represent the uncertainties coming from the error in the fitted 
parameters). The values of the parameters of Eqs.~(\ref{fit}) entering our 
results were obtained by a fit performed in the region 2.5 GeV$^2 \leq Q^2 \leq 20$ 
GeV$^2$, $10^{-4} \leq x \leq 5 
\times 10^{-3}$.} 
\label{fig1}
\end{figure}

Another test of this model regards the slope $dF_2^{\nu N} (x,Q^2)/d\ln Q^2$. 
As noticed in Ref.~\cite{JBA}, higher-twist effects are more easily 
revealed in the slope than in $F_2 (x,Q^2)$. Since in Ref.~\cite{JBA} it was
stated that one cannot rely on the DLLA for the calculation of the twist-4
contributions, a critical examination of our approximation becomes important.
It is easy to write in our approach (see Subsections~\ref{subsec:shadow}
and~\ref{subsec:small-x})
\begin{equation}
\frac{dF_2^{\nu N}(x,Q^2)}{d\ln Q^2}=\frac{df_q^{full}(x,Q^2)}{d\ln Q^2}
\sim \frac{df_q(x,Q^2)}{d\ln Q^2}+\frac{\alpha_s^2}{Q^2}K\left(-\frac{17}{32}
f_g^2(x,Q^2)\right).
\label{slope}
\end{equation}
In Figure~2 we show the behavior of the slope as a function of $x$ for $Q^2=10$ 
GeV$^2$ in the range $10^{-10} \leq x \leq 10^{-3}$ according to
our theoretical results, with the parameters given in Eqs.~(\ref{fit}). A similar
behavior is observed for other values of $Q^2$ in the range 2.5 GeV$^2 \leq 
Q^2 \leq 20$ GeV$^2$. From this Figure it is possible to see the higher-twist 
effects and to have the idea of the uncertainty resulting from the the errors
on the fitted parameters. 

\begin{figure}[tb]
\centering
\includegraphics[width=\textwidth]{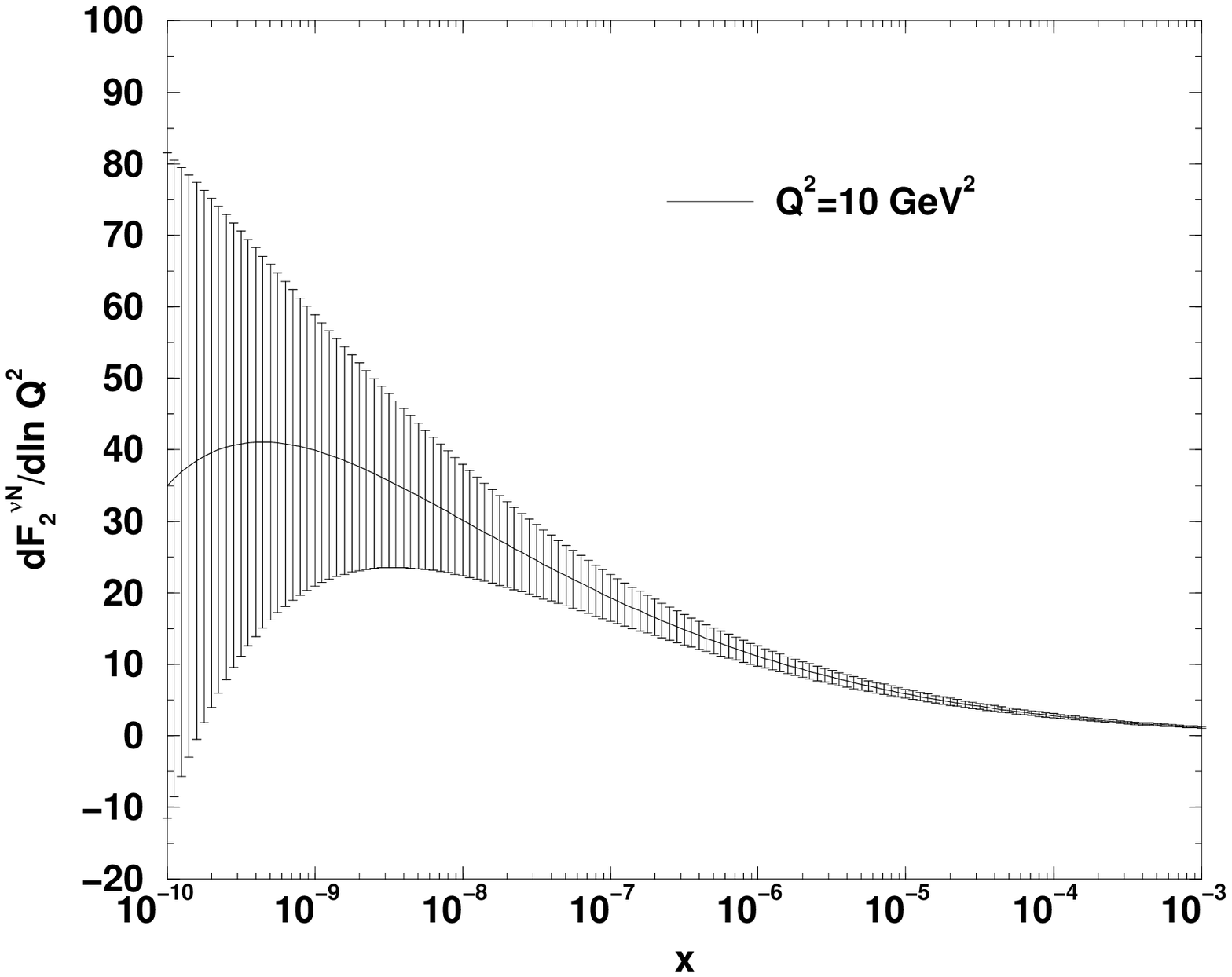}
\caption[]{\small Slope $dF_2^{\nu N}(x,Q^2)/d\ln Q^2$ at $Q^2=10$ GeV$^2$ 
according to our results. The uncertainties on the data come from the errors
in the values of the fitted parameters of Eqs.~(\ref{fit}).}
\label{fig2}
\end{figure}

Finally, we consider the total cross section. With the usual notation, we 
can write for charged-current neutrino interactions
\begin{eqnarray}
\!\!\!\!\! \left(\frac{d\sigma}{dx\,dy}\right)^{\nu} & = &
\frac{G_F^2ME_{\nu}}{\pi} \left(
\frac{M_W^2}{Q^2+M_W^2}\right)^2 \nonumber \\
&\times& \left\{\frac{1+(1-y)^2}{2}F_2^{\nu N}(x,Q^2)+
\left(1-\frac{y}{2}\right)yxF_3^{\nu N}(x,Q^2)\right\}\,. \label{nu2}
\end{eqnarray}
For anti-neutrino charged-current processes one must change the
sign in front of $F_3^{\nu N} (x,Q^2)$. In Eq.~(\ref{nu2}), $G_F$ is
the Fermi constant, $M$ is the nucleon mass and the variable $y$
is related to the Bjorken $x$ through the relation
\begin{equation}
y=\frac{Q^2}{x(s-M^2)}\simeq \frac{Q^2}{xs}\;. \label{nu3}
\end{equation}
The laboratory neutrino energy $E_{\nu}=(s-M^2)/(2M)$ is
approximately $s/(2M)$ in the energy region of interest. The $F_2$
contribution to  the total cross section can be written in the
form
\begin{displaymath}
\bar{\sigma}^{\nu N}\equiv\frac{\sigma^{\nu
N}+\sigma^{\bar{\nu}N}}{2}
\end{displaymath}
\begin{equation}
=\frac{G_F^2}{2\pi}\int_{Q_0^2}^s\,dQ^2\left(
\frac{M_W^2}{Q^2+M_W^2}\right)^2
\int_{Q^2/s}^{1}\,\frac{dx}{x}\;\frac{1+(1-Q^2/(xs))^2}{2}F_2^{\nu N}(x,Q^2)
\label{nu4}
\end{equation}
and the formulas obtained for the $F_2^{\nu N}(x,Q^2)$ allow
the evaluation of $\bar{\sigma}^{\nu N}$ that, for large $s$, is a
good approximation for the total $\nu N$ charged-current cross
section. 

In Figure~3 we show the behavior of $\bar{\sigma}^{\nu N}$ as a function
of $s$, in the range 10$^4$ GeV$^2 \leq s \leq 10^{14}$ GeV$^2$. The relative
error is 4.4\% at $s=10^4$ GeV$^2$ and 14.0\% at $s=10^{12}$ GeV$^2$.
For comparison we draw on the same plot the results obtained in Ref.~\cite{RG} 
and in Ref.~\cite{KMS} as well as the HERA measurement at $\sqrt s
= 306.42 $ GeV. At this value of $s$ our estimate of the cross section is 
much lower than the H1 data point. However, this is not surprising
in view of the fact that we neglected valence quarks, which still can play
a residual role at these relatively small values of $s$. For larger values of $s$, 
our results nicely compare with those of Refs.~\cite{RG,KMS}, showing good agreement
till $s \sim 10^{13}$ GeV$^2$. For even larger values of $s$ the effect of 
higher-twist starts to become visible. 

These findings support the conclusion that our approach, based on an
asymptotic calculation, succeeded in singling out the relevant part of 
the cross section.

In order to make the higher-twist effects more visible, we determined
also $\bar{\sigma}^{\nu N}$ without considering the recombination terms 
in the evolution equations, i.e. with $K=0$ from the beginning.
In this case the fitted parameters turn to be $A_q=1.040(36)$ and
$A_g=0.548(28)$ ($\chi^2$/d.o.f.=0.665). In Figure~4 we compare our results for
$\bar{\sigma}^{\nu N}$ with and without the inclusion of the recombination effect. 
The higher-twist effects become visible for $s=10^{13}$ GeV$^2$.

\begin{figure}[tb]
\centering
\includegraphics[width=\textwidth]{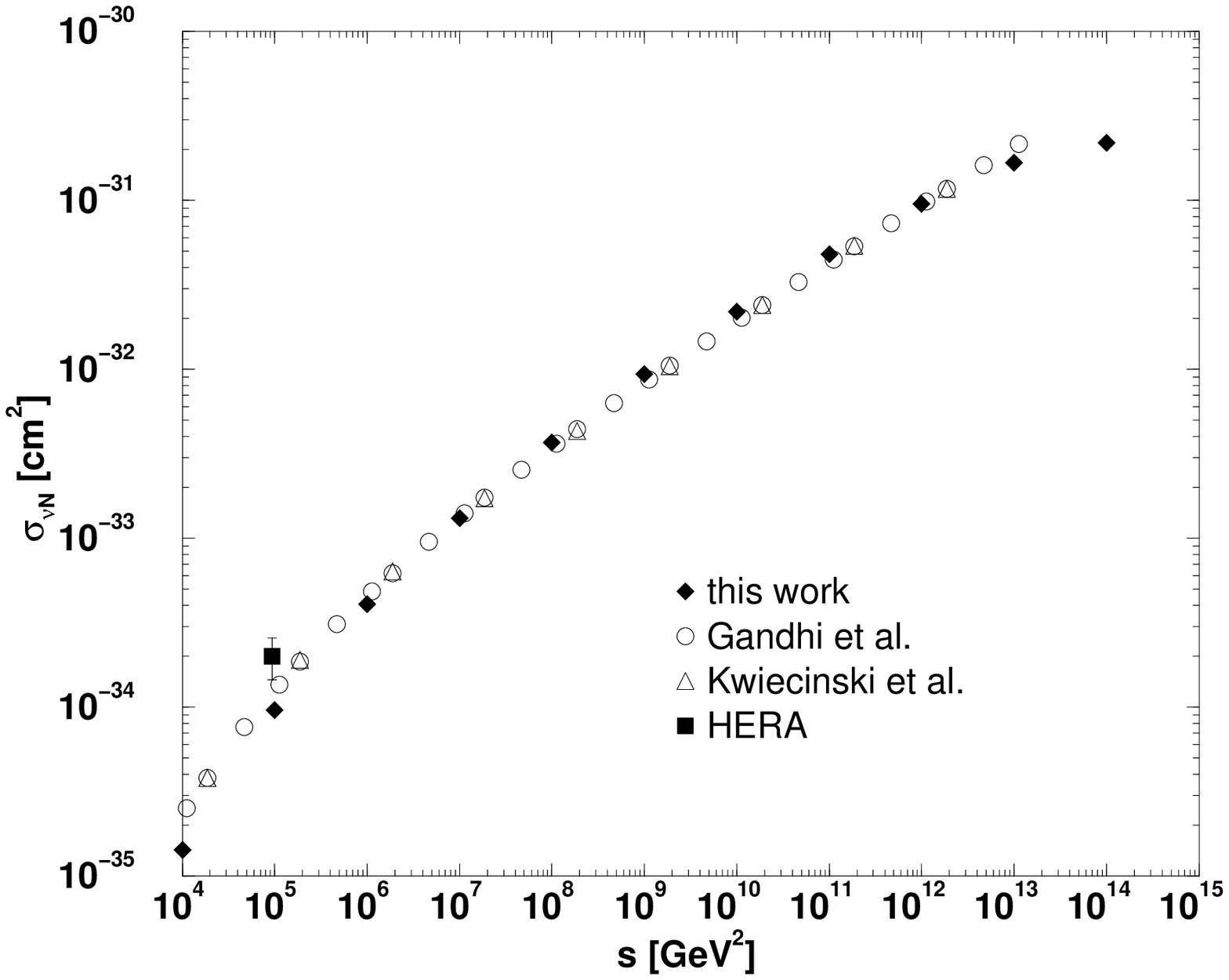}
\caption[]{\small Cross section $\bar{\sigma}^{\nu N}$ as a function of $s$
according to our results (filled diamonds). Data do not include the uncertainties 
coming from the errors on the values of the fitted parameters of Eqs.~(\ref{fit})
(see, instead, Figure~4). 
For comparison, data from Ref.~\cite{RG} (open circles) and from Ref.~\cite{KMS} 
(open triangles) are also shown. The isolated point at $\sqrt{s} = 306.42 $ GeV 
(filled squares) represents the HERA measurement.}
\label{fig3}
\end{figure}

\begin{figure}[tb]
\centering
\includegraphics[width=\textwidth]{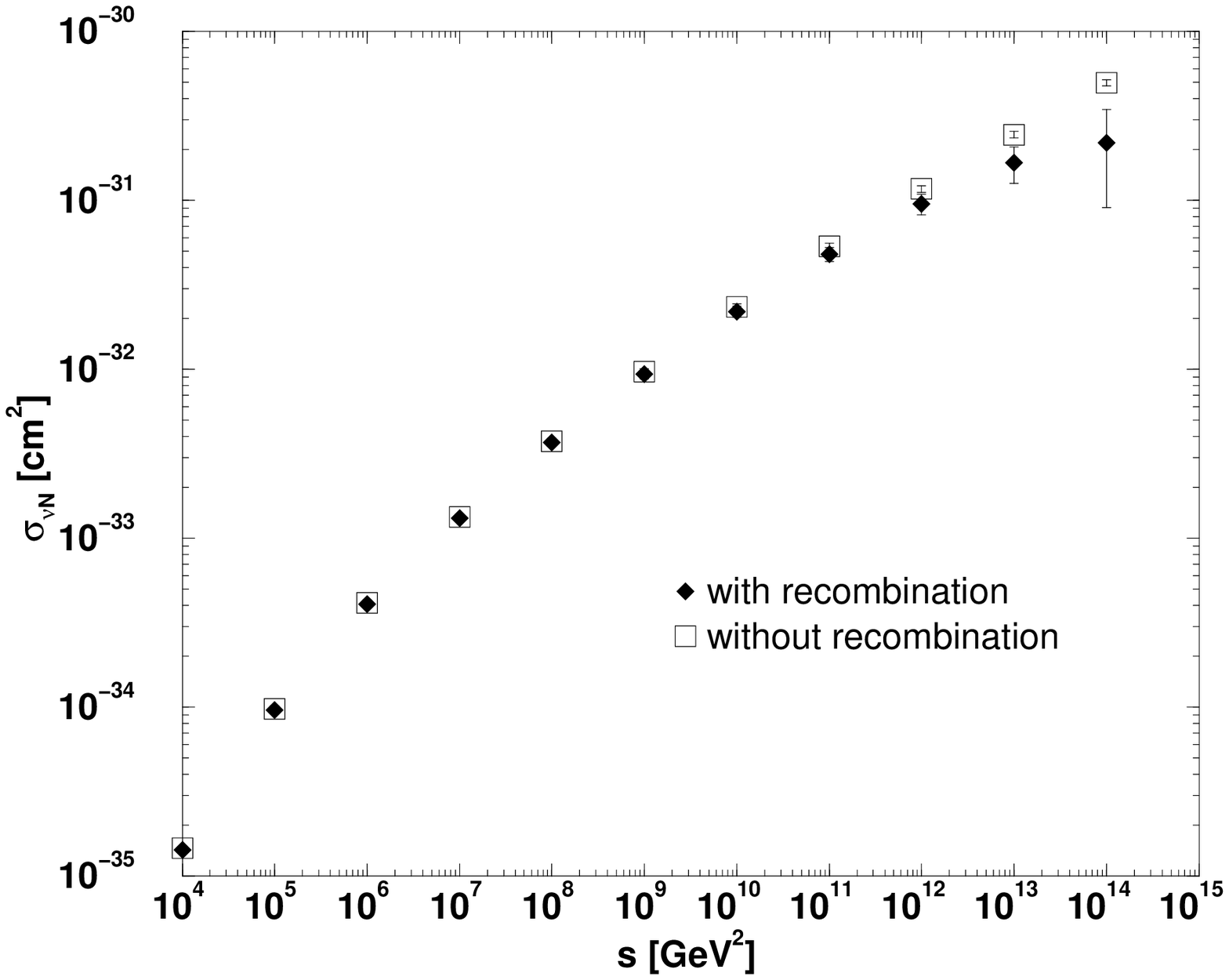}
\caption[]{\small Cross section $\bar{\sigma}^{\nu N}$ as a function of $s$
according to our results (filled diamonds, same as Figure~3, but with the inclusion
of error bars) and without the contribution from the recombination (open squares).}
\label{fig4}
\end{figure}

\section{Conclusions}
\label{sec:conclusions}

Twist-4 corrections to the structure function $F_2$ have been
estimated at small $x$ in leading order QCD following the method
developed in Ref.~\cite{AKP}. This estimate leads to an analytical
parametrization for the gluon recombination effects and completes,
in this respect, the program outlined in our previous
paper~\cite{FJKP}, where an approximate QCD evolution at twist-2
was presented. 

Our approach to the saturation phenomenon follows the scheme proposed 
in Ref.~\cite{WZ}. The non-linear evolution equations we study are the MD-DGLAP
equations~\cite{WZA}, where momentum conservation gives rise to an
antishadowing term that influences appreciably the screening
effects. A compact and analytical solution of these equations at
small $x$ is possible only if some conditions are satisfied. The
most relevant of these conditions regards the input parton
distributions that must be flat. The proof of Eq.~(\ref{e23}), 
which allows a simple treatment of the gluon recombination terms, requires 
this assumption. Moreover, the kernels in the MD-DGLAP equations are given 
in leading order of perturbation theory and this approximation reflects on our
approach. In spite of these assumptions and approximations, the
results are satisfactory. Examples of the $Q^2$ evolution and of
the behavior of the slope $d F_2/d \ln Q^2$, shown
in Figures~1 and 2, make us confident on the accuracy of the
method. 

The most interesting application of our simplified formulas is, in our opinion, 
the study of the interaction of ultra-high energy cosmic neutrinos with nucleons.
This process can probe an energy region far beyond the largest
energy reached by existing accelerators. The simplified and
reasonable expressions for the structure function $F_2^{\nu N}$,
disposable in this paper, renders the evaluation of the
neutrino-nucleon cross section and the estimate of the twist-4
contributions simpler and more transparent. Our result for the
cross section, shown in Figure~3, is in perfect agreement with the
calculations of other models and, only above $s=10^{13}$ GeV$^2$
($E_{\nu}\simeq 5.3\times 10^{12}$ GeV), the effect of twist-4
gluon recombination becomes visible. The Froissart limit will be
eventually satisfied, but in a region where the complete solution
for the gluon distribution becomes necessary. 

It is well possible, at such high energies, to simplify further the
integrals leading to the $\nu N$ cross section. Work on this
problem is in progress.

\vspace{0.5cm} {\bf \large Acknowledgments} L.J. and A.K. thank
the Departments of Physics of the Universities of Calabria and
Padova, together with the INFN Gruppo collegato di Cosenza and
Sezione di Padova, where this work was done, for their warm hospitality 
and support. The visit of A.K. to Padova was supported by INFN-LThPh agreement 
program.

\appendix

\section{Appendix: Proof of Eq.~(\ref{e23})}

We consider separately the three terms coming from the decomposition 
$f_g^2=(f_g^-)^2 + 2 f_g^- f_g^+ + (f_g^+)^2$, which will be named in the 
following ``$+ +$'', ``$+ -$'' and ``$- -$'' components, respectively. 
Moreover, we will use the notation $\stackrel{M}{\to}$ to denote
the Mellin transform and $\stackrel{M^{-1}}{\to}$ to denote the inverse 
Mellin transform. 

\noindent
{\bf a)} For the ``$- -$'' component (hereafter $A$ is arbitrary) we get
\begin{eqnarray}
 & & \int^{1/A}_x \, \frac{dy}{y} \, (f^-_g(y))^2 \sim
 \int^{1/A}_x \, \frac{dy}{y}=\ln \frac{1}{x}-\ln A \nonumber \\
& \sim &  \left( \ln \frac{1}{x}-\ln A \right) (f^-_g(x))^2
~\stackrel{M}{\to} ~ \left( \frac{1}{n-1}-\ln A \right)
f^{--}_2(n)\;, \label{a1}
\end{eqnarray}
where we have used the definition
\begin{equation}
f_2^{ij}(n)= \int^{1}_0 \, dx \, x^{n-2} \, f^i_g(x) f^j_g(x)~~~
(i,j=\pm) \;. \label{a2}
\end{equation}
The same argument gives
\begin{eqnarray}
 & & \int^{1/A}_{x/2} \, \frac{dy}{y} \,(f^-_g(y))^2 \sim
 \int^{1/A}_{x/2} \, \frac{dy}{y}=\ln \frac{2}{x}-\ln A \nonumber \\
&\sim & \left( \ln \frac{2}{x}-\ln A \right) (f^-_g(x))^2
~\stackrel{M}{\to} ~ \left( \frac{1}{n-1}+\ln 2 -\ln A \right)
f^{--}_2(n)\,. \label{a3}
\end{eqnarray}
Then, from Eqs.~(\ref{a1}) and (\ref{a3}) we obtain
\begin{equation}
\int^{x}_{x/2} \, \frac{dy}{y} \, (f^-_g(y))^2 = \ln 2 \,\,
(f^-_g(x))^2 ~\stackrel{M}{\to} ~ \ln 2 \,\, f^{--}_2(n)\;.
\label{a4}
\end{equation}

\noindent {\bf b)} For the ``$+ -$'' component
(hereafter $z=\ln(1/y)$ and $\Delta=|\hat d_{gg}|t$) we get
\begin{eqnarray}
&& \int^{1/A}_x \,\frac{dy}{y} \, f^+_g(y) f^-_g(y) ~\sim~
 \int^{1/A}_x \, \frac{dy}{y} \, I_0(\sigma (y)) =
\int^{\ln(1/x)}_{\ln A} \, dz \, \sum_{k=0}^{\infty}
\frac{\Delta^k z^k}{(k!)^2} \nonumber \\
&& = \sum_{k=0}^{\infty} \frac{\Delta^k}{k! (k+1)!} \left[
\left(\ln \frac{1}{x}\right)^{k+1} -\left(\ln A\right)^{k+1}
\right]
= \frac{1}{\rho} I_1(\sigma) - \frac{1}{\rho_A} I_1(\sigma_A) \nonumber \\
&& =\frac{1}{\rho} I_1(\sigma)\cdot \Bigl(1 + O(\rho^2)\Bigr)
~\stackrel{M}{\to} ~ \left( \frac{1}{n-1} + O(n-1)\right)
f^{+-}_2(n)\;, \label{a5}
\end{eqnarray}
where
\begin{equation}
\rho_A= \rho |_{x\to 1/A},~~~ \sigma_A = \sigma |_{x\to 1/A}\;.
\label{a6}
\end{equation}
Similarly, we obtain
\begin{equation}
\int^{1/A}_{x/2} \, \frac{dy}{y} \, f^+_g(y) f^-_g(y) ~\sim~
\frac{1}{\rho_2} I_1(\sigma_2) - \frac{1}{\rho_A} I_1(\sigma_A)\;,
\label{a7}
\end{equation}
where
\begin{equation}
\rho_2= \rho |_{x\to x/2}, \hspace{1cm} \sigma_2= \sigma |_{x\to x/2}\;.
\label{a8}
\end{equation}
Note that
\begin{eqnarray}
&& \frac{1}{\rho_2} I_1(\sigma_2)= \sum_{k=0}^{\infty}
\frac{\Delta^k}{k! (k+1)!} \left(\ln \frac{2}{x}\right)^{k+1}
\nonumber \\
&& =\sum_{k=0}^{\infty} \frac{\Delta^k}{k! (k+1)!} \left[
\left(\ln \frac{1}{x}\right)^{k+1} + \! (k+1) \ln 2 \left(\ln
\frac{1}{x}\right)^{k}
+ O \left(\left(\ln \frac{1}{x}\right)^{k-1}\right) \right] \nonumber \\
&& =\frac{1}{\rho} I_1(\sigma) +\ln 2 I_0(\sigma) \ \cdot \Bigl(1
+ O(\rho)\Bigr) \;. \label{a9}
\end{eqnarray}
Then we arrive at
\begin{eqnarray}
\int^{1/A}_{x/2} \, \frac{dy}{y} \, f^+_g(y) f^-_g(y) &\sim&
\frac{1}{\rho} I_1(\sigma) +\ln 2 I_0(\sigma) \
\cdot \Bigl(1 + O(\rho)\Bigr) \nonumber \\
&\stackrel{M}{\to} & \left( \frac{1}{n-1} +\ln 2 + O(n-1)\right)
f^{+-}_2(n) \,.\label{a10a}
\end{eqnarray}
Finally, from Eqs.~(\ref{a5}) and (\ref{a10a}), we obtain
\begin{eqnarray}
\int^{x}_{x/2} \, \frac{dy}{y} \, f^+_g(y) f^-_g(y) &=& \ln 2 \,\,
f^+_g(x) f^-_g(x)\cdot \Bigl(1 + O(\rho)\Bigr) \nonumber \\
 &\stackrel{M}{\to}& \Bigl( \ln 2 + O(n-1) \Bigr)\,\, f^{+-}_2(n)\;.
\label{a10}
\end{eqnarray} 

\noindent {\bf c)} For the ``$+ +$'' component we get
\begin{eqnarray}
&& \int^{1/A}_x \, \frac{dy}{y} \, (f^+_g(y))^2 \sim
 \int^{1/A}_x \, \frac{dy}{y} \, I_0^2(\sigma (y))
= \int^{\ln(1/x)}_{\ln A} \, dz \, I_0^2(\sigma (z))
\nonumber \\
&& = \frac{1}{\rho} I_1(\sigma) I_0(\sigma) \cdot \Bigl(1 +
O(\rho^2)\Bigr) - \int^{\ln(1/x)}_{\ln A} \, dz \, I_1^2(\sigma
(z))\;, \label{a11}
\end{eqnarray}
where we have used the integration by parts and the results from the
previous case {\bf b)}.

Since, for $\sigma \to \infty$, $I_1(\sigma) = I_0(\sigma) \cdot
(1 + O(\rho))$, we have
\begin{eqnarray}
\int^{1/A}_x \,\frac{dy}{y} \, (f^+_g(y))^2 & \sim &
\frac{1}{2\rho} I_1(\sigma) I_0(\sigma) \cdot \Bigl(1 +
O(\rho^2)\Bigr)
\nonumber \\
&\stackrel{M}{\to} & ~ \left( \frac{1}{n-1} + O(n-1)\right)
f^{++}_2(n) \;. \label{a12}
\end{eqnarray}
In the same way, we obtain
\begin{eqnarray}
\int^{1/A}_{x/2} \, \frac{dy}{y} \, (f^+_g(y))^2 &\sim &
\frac{1}{2\rho_2} I_1(\sigma_2) I_0(\sigma_2)
\cdot \Bigl(1 + O(\rho^2)\Bigr) \nonumber \\
&=& \frac{1}{2\rho} I_1(\sigma)I_0(\sigma) +\ln 2 \, \,
I^2_0(\sigma) \
\cdot \Bigl(1 + O(\rho)\Bigr) \nonumber \\
&\stackrel{M}{\to} & \left( \frac{1}{n-1} +\ln 2 + O(n-1)\right)
f^{++}_2(n) \label{a13}
\end{eqnarray}
and, from Eqs.~(\ref{a12}) and (\ref{a13}),
\begin{eqnarray}
\int^{x}_{x/2} \, \frac{dy}{y} \, (f^+_g(y))^2 &=& \ln 2 \,\,
(f^+_g)^2 \cdot \Bigl(1 + O(\rho)\Bigr) \nonumber \\
&\stackrel{M}{\to}& \Bigl( \ln 2 + O(n-1) \Bigr)\,\, f^{++}_2(n)\;.
\label{a14}
\end{eqnarray}
Finally we get
\begin{eqnarray}
\int^{1/A}_x \, \frac{dy}{y} \, f^2_g(y) &=& \int^{1}_x \,
\frac{dy}{y} \, f^2_g(y) \cdot \Bigl(1 + O(\rho^2)\Bigr)
-\ln A \, (f^-_g(x))^2 \nonumber \\
&=& \int^{1}_x \, \frac{dy}{y} \, f^2_g(y) \cdot \Bigl(1 +
O(\rho^2)\Bigr) \nonumber \\
& \stackrel{M}{\to} & \left( \frac{1}{n-1} +
O(n-1)\right) f_2(n)\;, \label{a15}
\end{eqnarray}
because (see Eqs.~(\ref{u5}) and (\ref{u7})) 
\begin{equation}
\frac{(f^-_g(x))^2}{f^2_g(x)} \, \sim \,I_0^{-2}(\sigma) \ll
O(\rho^2)\;. \label{a16}
\end{equation}
Thus, the result of~(\ref{a15}) does not depend on the specific
value of $A$. Analogously we find
\begin{eqnarray}
\int^{1/A}_{x/2} \, \frac{dy}{y} \, f^2_g(y) &=& \int^{1}_x \,
\frac{dy}{y} \, f^2_g(y) + \ln 2 \,\, f^2_g(x)
\cdot \Bigl(1 + O(\rho)\Bigr) \nonumber \\
&\stackrel{M}{\to} & \left( \frac{1}{n-1} +\ln 2 + O(n-1)\right)
f_2(n) \;.\label{a17}
\end{eqnarray}
By comparing Eq.~(\ref{a17}) for $A$=2 and Eq.~(\ref{e20}), we get
\begin{equation}
2^{n-1}\, \frac{1}{n-1}\,f_2(n|1/2) = \left( \frac{1}{n-1} +\ln
2 + O(n-1)\right) f_2(n) \label{a18}
\end{equation}
and conclude that
\begin{equation}
f_2(n|1/2) = f_2(n) + O((n-1)^2) \;.\label{a19}
\end{equation}
This completes the proof of Eq.~(\ref{e23}). Equation~(\ref{a19}) 
is very important and helps us to sum the regular parts of $F_{ag}(x)$. 
We should remember, however, that this result
holds only for $x$-independent input distributions, that is for
$B_q=B_g=0$ in Eq.~(\ref{e2}).

Comparing Eqs.~(\ref{a4}), (\ref{a10}) and (\ref{a14}), we get finally
\begin{equation}
\int^{x}_{x/2} \, \frac{dy}{y} \, f^2_g(y) =  \ln 2 \,\,
f^2_g(x) \cdot \Bigl(1 + O(\rho)\Bigr) ~\stackrel{M}{\to} ~ \ln 2
\,\, f_2(n) + O(n-1)\;. \label{a20}
\end{equation}

\section{Appendix: The coefficients $C_{ag}^{i,\pm}$}

Here we give the explicit values of the coefficients
$C_{ag}^{i,\pm}(n=1) \equiv C_{ag}^{i,\pm}$ appearing in 
Eqs.~(\ref{e32}) and (\ref{e33}):
\begin{eqnarray}
&& C_{qg}^{1,-}=\frac{283}{2880}\;, \hspace{1cm}
C_{qg}^{2,-} =  \frac{1813}{2880} - \frac{297}{32} f\;, \nonumber \\
&& C_{qg}^{1,+} =  0\;, \hspace{1.7cm}
C_{qg}^{2,+} = \frac{297}{32} f\;, \nonumber \\
&& C_{gg}^{1,-} = - \frac{283}{6480}\;, \hspace{0.7cm} 
C_{gg}^{2,-} =  \frac{33}{8} f - \frac{1813}{6480}\;, \nonumber \\
&& C_{gg}^{1,+} =  \frac{27}{32} \left(99 \ln 2 -
\frac{255361}{4374} \right) \;. \label{a21}
\end{eqnarray}

For the coefficient $C_{gg}^{2,+}$ one must consider separately
the singular and regular part,
\begin{displaymath}
 C_{gg}^{2,+} = \hat C_{gg}^{2,+} \, \frac{1}{n-1}
+ \overline C_{gg}^{2,+}(n=1)\;,
\end{displaymath}
where
\begin{equation}
\hat C_{gg}^{2,+} = \frac{2673}{32}\;, \hspace{1cm} 
\overline C_{gg}^{2,+}(n=1) =  - \frac{33}{8}f - \frac{516577}{5184} \;.
\label{a22}
\end{equation}

\section{Appendix: Proof of the simplified form for the functions ${\tilde{F_1}}$ 
and ${\tilde{F_2}}$}

The validity of the simplified form for the functions
${\tilde{F_1}}$ and ${\tilde{F_2}}$, given in Eqs.~(\ref{e52})-(\ref{e57}), 
is based on the following estimates. As in Appendix~A, we consider
separately the three components of the functions ${\tilde{F_1}}$ and 
${\tilde{F_2}}$ given in Eq.~(\ref{e51}).

\noindent {\bf a)} Consider first the ``$- -$'' component. 
The function $(f^{-}_g(x,M^2))^2$
is $x$-independent since we assumed $B_q=B_g=0$ :
$(f^{-}_g(x,M^2))^2=(f^{-}_g(M^2))^2$, so
\begin{equation}
f_2^{--}(n,M^2) = \frac{1}{n-1} \, (f^{-}_g(M^2))^2 \;.
 \label{a23}
\end{equation}
Then (see Eqs.~(\ref{e38}) and (\ref{e39})) we have
\begin{eqnarray}
\Bigl[\delta(1-x) - {\tilde{\rho}} J_1({\tilde{\sigma}}) \Bigl]
&\otimes & (f^{-}_g(x,M^2))^2 ~~\stackrel{M}{\to} ~~
e^{\hat d_+v/(n-1)} \, \frac{1}{n-1} \, (f^{-}_g(M^2))^2 \nonumber \\
&&~~\stackrel{M^{-1}}{\to} ~~
J_0({\tilde{\sigma}}) (f^{-}_g(x,M^2))^2 \;, \label{a24} \\
J_0({\tilde{\sigma}}) &\otimes & (f^{-}_g(x,M^2))^2
~~\stackrel{M}{\to} ~~
e^{\hat d_+v/(n-1)} \, \frac{1}{(n-1)^2} \, (f^{-}_g(M^2))^2 \nonumber \\
&&~~\stackrel{M^{-1}}{\to} ~~ \frac{1}{{\tilde{\rho}}}\,
J_1({\tilde{\sigma}}) (f^{-}_g(x,M^2))^2  \;. \label{a25}
\end{eqnarray}
Thus,
\begin{eqnarray}
\tilde F^{--}_1 &=& (A_g^-)^2 J_0({\tilde{\sigma}}) \cdot e^{-
2d_{-}(1) w} \Bigl(1~+~O(\rho)\Bigr)\;,
 \label{a26}\\
\tilde F^{--}_2 &=& (A_g^-)^2 \frac{1}{\tilde \rho}
J_1({\tilde{\sigma}}) \cdot e^{- 2d_{-}(1) w}
\Bigl(1~+~O(\rho)\Bigr)\;.
 \label{a27}
\end{eqnarray}

\noindent {\bf b)} Consider now the ``$+ -$'' component. 
The function $f^{+}_g(x,M^2)$ can be represented as
\begin{equation}
f^{+}_g(x,M^2)=A_g^+ \, I_0({\hat{\sigma}})\, e^{-\overline d_+(1)w}
 ~~\stackrel{M}{\to} ~~\frac{1}{n-1} \, A_g^+ \,
e^{-\hat d_+w/(n-1)} \, e^{-\overline d_+(1)w} \label{a28} \;.
\end{equation}
Note that
\begin{eqnarray}
\Bigl[\delta(1-x) - {\tilde{\rho}}J_1({\tilde{\sigma}}) \Bigl]
 &\otimes & I_0({\hat{\sigma}})
~~\stackrel{M}{\to}~~ e^{\hat d_+v/(n-1)} \, \frac{1}{n-1} \,
e^{-\hat d_+w/(n-1)} \nonumber \\
&=& \frac{1}{n-1} \, e^{-\hat d_+t/(n-1)}
~~\stackrel{M^{-1}}{\to}~~ I_0(\sigma) \;. \label{a29}
\end{eqnarray}
In the same way we obtain
\begin{eqnarray}
J_0({\tilde{\sigma}}) &\otimes & I_0({\hat{\sigma}})
~~\stackrel{M}{\to}~~ \frac{1}{n-1} \, e^{\hat d_+v/(n-1)} \,
\frac{1}{n-1} \, e^{-\hat d_+w/(n-1)}
\nonumber \\
&=& \frac{1}{(n-1)^2} \, e^{-\hat d_+t/(n-1)}
~~\stackrel{M^{-1}}{\to}~~ \frac{1}{\rho}\, I_1(\sigma) \;.
\label{a30}
\end{eqnarray}
Thus, we get
\begin{eqnarray}
\tilde F^{+-}_1 &\equiv& - {\tilde{\rho}} J_1({\tilde{\sigma}})
\otimes  \Bigl(f^{+}_g(x,M^2) f^{-}_g(x,M^2)\Bigr) \nonumber \\
&=& A_g^- A_g^{+} I_0(\sigma) \cdot e^{- (d_{-}(1)+\overline
d_{+}(1)) w} \Bigl(1~+~O(\rho)\Bigr)\;,
 \label{a31}\\
\tilde F^{+-}_2 &\equiv& J_0({\tilde{\sigma}})
\otimes  \Bigl(f^{+}_g(x,M^2)f^{-}_g(x,M^2)\Bigr) \nonumber \\
&=& A_g^- A_g^{+} \frac{1}{\rho} I_1(\sigma) \cdot e^{-
(d_{-}(1)+\overline d_{+}(1)) w} \Bigl(1~+~O(\rho^2)\Bigr) \;.
 \label{a32}
\end{eqnarray}

\noindent {\bf c)} Finally, let us consider the ``$+ +$'' component. 
The problem of finding the small-$x$ behavior
of $J_0(\tilde{\sigma})\otimes I_0^2(\hat{\sigma})$ can be solved
as follows. From the tables of the inverse Mellin transforms in Ref.~\cite{BAT}, we 
obtain
\begin{displaymath}
\int_0^1\,I_{\nu}[(\alpha^{1/2}+\beta^{1/2})y^{1/2}]I_{\nu}[(\alpha^{1/2}-
\beta^{1/2})y^{1/2}]\,x^{\gamma-1}\,dx
\end{displaymath}
\begin{equation}
=\frac{1}{\gamma}\,\exp\left(\frac{\alpha+\beta}{2
\gamma}\right)\,I_{\nu}\left(\frac{\alpha-\beta}{2\gamma}\right)\;,
\label{a33}
\end{equation}
where $y=-\log x$ and Re $\nu >-1$, Re $\gamma>0$. 

Equation (\ref{a33}) gives, for $\beta=\nu=0,\;\alpha=-4
\hat{d}_{gg}w$ and $\gamma=n-1$, the Mellin transform of
$I_0^2(\hat{\sigma})$:
\begin{equation}
I_0^2(\hat{\sigma})~~\stackrel{M}{\to}~~\frac{1}{n-1}\,e^{-2\hat{d}_+w/(n-1)}
I_0\left(\frac{-2\hat{d}_+w}{n-1} \right) \;.\label{a34}
\end{equation}
The same result can be obtained in an alternative way by performing the 
Mellin transform of the series
\begin{displaymath}
I_0^2(\hat{\sigma})=\frac{1}{\sqrt{\pi}}\sum_0^{\infty}
\frac{(\hat{\sigma})^{2m} \Gamma(m+1/2)}{\Gamma^2(m+1)} \;.
\end{displaymath}
From this first step we get
\begin{equation}
J_0(\tilde{\sigma})\otimes
I_0^2(\hat{\sigma}) ~~\stackrel{M}{\to}~~
\frac{1}{(n-1)^2}\,e^{-\hat{d}_+(w+t)/(n-1)} I_0\left(
\frac{-2\hat{d}_+w}{n-1} \right) \label{a35}
\end{equation}
and the inverse Mellin transform of the right hand side of (\ref{a35}) 
will be the final answer for $\tilde F^{++}_2$ (apart from the overall factor
$(A_g^+)^2$).

We consider now Eq.~(\ref{a33}) for $\nu=0$. By taking the sum of the 
derivative of Eq.~(\ref{a33}) with respect to $\alpha$ 
with its derivative with respect to $\beta$, we obtain the important formula
\begin{eqnarray}
&& \frac{1}{2\sqrt{\alpha\beta}}\,\int_0^1\,dx x^{\gamma-1}
y^{1/2} \nonumber \\ 
&& \times \left[(-\alpha^{1/2}+\beta^{1/2})I_0[(\alpha^{1/2}+\beta^{1/2})y^{1/2}]
I_1[(\alpha^{1/2}-\beta^{1/2})y^{1/2}]+ \right. \nonumber \\ &&
\left. +(\alpha^{1/2}+\beta^{1/2})I_0[(\alpha^{1/2}-\beta^{1/2})
y^{1/2}]\,I_1[(\alpha^{1/2}+\beta^{1/2})y^{1/2}]\right] \nonumber
\\ &&  = \frac{1}{\gamma^2}\,e^{(\alpha+\beta)/(2 \gamma)}I_0\left(
\frac{\alpha-\beta}{2 \gamma} \right) \;, \label{a36}
\end{eqnarray}
where, as before, $\gamma\equiv n-1$ and $y=-\log x$. With the
substitutions
\begin{equation}
\alpha \rightarrow -\hat{d}_+ (3 w+t),
\hspace{1cm} \beta \rightarrow -\hat{d}_+(t-w) \;,\label{a37}
\end{equation}
Eq.~(\ref{a36}) provides the final answer, since the
coefficient of $x^{\gamma-1}$ in the integral at the left hand
side gives the required inverse Mellin transform. 

A similar procedure can be applied to determine $\tilde{\rho} J_1(\tilde{\sigma})
\otimes I_0^2(\hat{\sigma})$, which is needed to calculate
$\tilde F^{++}_1$, starting from the formula~(\ref{a33}), with the 
same values of the parameters $\alpha$ and $\beta$ as in~(\ref{a37})
and $y=-\log x$. Notice that $\beta<0$ and hence $\beta^{1/2}$ is pure imaginary. 

After introducing, for the sake of simplicity, the notation 
$z=\alpha^{1/2}+i \left|\beta^{1/2} \right|$, we get
 \begin{eqnarray}
\tilde F^{++}_1 &\equiv& \Bigl[\delta(1-x) - {\tilde{\rho}}
J_1({\tilde{\sigma}}) \Bigr]
\otimes  \Bigl(f^{+}_g(x,M^2)\Bigr)^2 \nonumber \\
&=& ( A_g^{+})^2 \,I_0(zy^{1/2})I_0(\bar{z}y^{1/2}) \cdot e^{-2
\overline d_{+}(1) w}
 \cdot
\Bigl(1~+~O(\rho)\Bigr)\;, \label{a38}
\end{eqnarray}
while, with the same definition of the variables $\alpha$, $\beta$ and
$y$, we get
\begin{eqnarray}
\tilde F^{++}_2 &\equiv& J_0({\tilde{\sigma}})
\otimes  \Bigl(f^{+}_g(x,M^2)\Bigr)^2 \nonumber \\
&=& ( A_g^{+})^2 \,\frac{1}{2\sqrt{\alpha\beta}} y^{1/2}\left[ z
I_0(\bar{z}y^{1/2})I_1(zy^{1/2}) \right. \nonumber \\ & - & \left.
\bar{z} I_0(z y^{1/2}) I_1(\bar{z} y^{1/2}) \right] \cdot e^{-2
\overline d_{+}(1) w} \cdot \Bigl(1~+~O(\rho^2)\Bigr)\;. \label{a39}
\end{eqnarray}

This completes the calculation of the simplified form of the functions 
${\tilde{F_1}}$ and ${\tilde{F_2}}$.

\vfill \eject

\end{document}